\documentclass[letterpaper,11pt,twocolumn]{./IEEEtran11}
\usepackage{graphicx}

\begin{document}

\title{SAFIUS - A secure and accountable filesystem over untrusted storage}
\author{V Sriram, Ganesh Narayan, K Gopinath\\
Computer Science and Automation\\
Indian Institute of Science, Bangalore\\
\{v\_sriram, nganesh, gopi\}@csa.iisc.ernet.in
}
\maketitle

\begin{abstract}
We describe SAFIUS, a secure accountable file system that resides over
an untrusted storage. SAFIUS provides strong security guarantees like
confidentiality, integrity, prevention from rollback attacks, and
accountability. SAFIUS also enables read/write sharing of data and provides the
standard UNIX-like interface for applications. To achieve accountability with
good performance, it uses asynchronous signatures; to reduce the space required
for storing these signatures, a novel signature pruning mechanism is used.
SAFIUS has been implemented on a GNU/Linux based system modifying OpenGFS.
Preliminary performance studies show that SAFIUS has a tolerable overhead for
providing secure storage: while it has an overhead of about  50\% of OpenGFS in
data intensive workloads (due to the overhead of performing
encryption/decryption in software), it is comparable (or better in some cases)
to OpenGFS in metadata intensive workloads.
\end{abstract}

\section{Introduction}
\label{cha:intro}
With storage requirements growing at around 40\% every year, deploying and
managing enterprise  storage is becoming increasingly problematic. The need for
ubiquitous storage accessibility also requires a re-look at traditional
storage architectures. Organizations respond to such needs by centralizing the storage
management: either inside the organization, or by outsourcing the storage.
Though both the options are together feasible and can coexist, they both pose
serious security hazards: the user can no longer afford to implicitly trust the
storage or the storage provider/personnel with critical data.

Most systems respond to such a threat by protecting data cryptographically
ensuring confidentiality and integrity. However, conventional security measures
like confidentiality and update integrity alone are not sufficient in managing
long lived storage: the storage usage needs to be accounted, both in quality and
quantity; also the inappropriate accesses, as specified by the user, should be
disallowed and individual accesses should ensure non-repudiation. In order for
such storage to be useful, the storage accesses should also provide freshness
guarantees for updates.

In this work we show that it is possible to architect such a secure and
accountable file system over an untrusted storage which is administrated in
situ or outsourced. We call this architect SAFIUS: SAFIUS is designed to
leverage trust onto an easily manageable entities, providing secure
access to data residing on untrusted storage. The critical aspect of SAFIUS
that differentiates it from rest of the solutions is that storage clients
themselves are independently managed and need not mutually trust each other.

\subsection{Data is mine, control is not!}
\label{sec:sysadm}
In many enterprise setups, users of data are different from the ones who
control the data: data is managed by storage administrators, who are neither
producers nor consumers of the data. This requires the users to trust storage
administrators without an option. Increased storage requirements could result
in an increase in the number of storage administrators and users would be
forced to trust a larger number of administrators for their data. A survey\footnote{http://www.storagetek.com.au/company/press/survey.html}, by
Storagetek, revealed that storage administration was a major cause of
difficulty in storage management as data storage requirements increased.
Although outsourcing of storage requirements is currently small, with continued
explosion in the data storage requirements and sophistication of technologies
needed to make the storage efficient and secure, enterprises may soon outsource
their storage (management) for cost and efficiency reasons. Storage service
providers (SSPs) provide storage and its management as a service. Using
outsourced storage or storage services would mean that entities outside an
enterprise have access to (and in fact control) enterprise's data.

\subsection{Need to treat storage as an untrusted entity}
Hence, there is a strong need to treat storage as an untrusted entity.
Systems like PFS\cite{pfs}, Ivy\cite{ivy}, SUNDR\cite{sundr},
Plutus\cite{plutus} and TDB\cite{tdb} provide a secure filesystem over an
untrusted storage. Such a secure filesystem needs to provide integrity
and confidentiality guarantees. But, that alone is not sufficient as the server
can still disseminate old, but valid data to the users in place of the most
recent data (rollback attack \cite{plutus}). Further, the server, if malicious,
cannot be trusted to enforce any protection mechanisms (access control) to
prevent one user from dabbling with another user's data which he is not
authorized to access. Hence a malicious user in collusion with the server can
mount a number of attacks on the system unless prevented.

All systems mentioned above protect the clients from the servers. But, we
argue that \emph{we also need to protect the server from malicious clients}. If
we do not do this, we may end up in a situation where the untrusted storage
server gets penalized even when it is not malicious.  If the system allows
arbitrary clients to access the storage, then it would be difficult to control
each of these clients to obey the protocol. The clients themselves could be
compromised or the users who use the clients could be malicious. Either way the
untrusted storage server could be wrongly penalized. To our knowledge, most
systems implicitly trust the clients and may not be useful in certain
situations.

\subsection{SAFIUS - Secure Accountable filesystem over untrusted storage}
We propose SAFIUS, an architecture that provides 
accountability guarantees apart from providing secure access to data
residing on untrusted storage.  By leveraging on an easily manageable
trusted entity in the system, we provide secure access to a scalable
amount of data (that resides on an untrusted storage) for a number of
\emph{independently managed} clients. The trusted entity is needed
only for maintaining some global state to be shared by many clients;
the bulk data path does not involve the trusted entity.    
SAFIUS guarantees that a party that violates the security protocol can
\emph{always} be identified precisely, preventing entities which obey the
protocol from getting penalized. The party can be one of the clients which
exports filesystem interface to users or the untrusted storage.

The following are the high level features of SAFIUS

\begin{itemize}
\item It provides confidentiality, integrity and freshness guarantees
  on the data stored.
\item It can identify the entities that violate the protocol.
\item It provides sharing of data for reading and writing among users.
\item Clients can recover independently from failures without
  affecting global filesystem consistency.
\item It provides close to UNIX like semantics.
\end{itemize}  

The architecture is implemented in GNU/Linux. Our studies show that SAFIUS has
a tolerable overhead for providing secure storage: while it has an overhead of
about 50\% of OpenGFS for data intensive workloads, it is comparable (or better
in some cases) to OpenGFS in metadata intensive workloads.

\section{Design}
\label{sec:design}
SAFIUS provides secure access to data stored on an
untrusted storage with perfect accountability guarantees by
maintaining some global state in a trusted entity.
In the SAFIUS system, there are \emph{fileservers}\footnote{They are termed fileservers as
these machines can potentially serve as NFS servers with a looser
consistency semantics to end clients.}
that provide filesystem access to clients,
with the back-end storage residing on \emph{untrusted storage}, henceforth
referred to as \emph{storage server}. The \emph{fileservers} can reach the
\emph{storage servers} directly. The system also has lock servers, known as
\emph{l-hash server} (for lock-hash server), a trusted entity that provides locking service and 
also holds and serves some critical metadata. 

\subsection{Security requirements}
Since the filesystem is built over an untrusted data store, it is
mandatory to have confidentiality, integrity and
freshness guarantees for the data stored. These guarantees
prevent the exposure or update of data by the storage server either by
unauthorized users or by collusion between unauthorized users and the
storage server.  Wherever there is mutual distrust between
entities, protocols employed by the system should be able to identify
the misbehaving entity (entity which violates the protocol)
precisely. This feature is referred to as \emph{accountability}.

\subsection{Sharing and Scalability}
The system should enable easy and seamless sharing of data between
users in a safe way. Users should be able to modify sharing semantics
of a file on their own, without the involvement of a trusted entity. 
The system should also be scalable to a
reasonably large number of users.

\subsection{Failures and recoverability}
The system should continue to function, tolerating failures of the
fileservers and it should be able to recover from failures of l-hash
servers or storage servers. Fileservers and storage servers can fail
in a byzantine manner as they are not trusted and hence can be malicious.
The fileservers should recover independently from failures.

\subsection{Threat Model}
SAFIUS is based on a relaxed threat model:

\begin{itemize}
  
\item Users \emph{need not} trust all the fileservers uniformly.  They
  need to trust only those fileservers through which they access the
  filesystem. Even this trust is temporal and can be revoked. It is
  quite impractical to build a system without the user to fileserver
  trust\footnote{The applications which access the data would not have any
  assurance on the data read or written as it passes through an
  untrusted operating system}.
  
\item No entity trusts the storage server and vice versa. The storage
  server is not even trusted for correctly storing of data.
  
\item The users and hence the fileservers need not trust each other
  and we assume that they do not. This assumption is important for
  ease of management of the fileservers. The fileservers can be
  independently managed and the users have the choice and
  responsibility to choose which fileservers to trust.
  
\item The l-hash servers are trusted by the fileservers, but
  \emph{not} vice versa.

\end{itemize}

\begin{figure}[t]
\includegraphics [width=3.25in, height=1.5in]{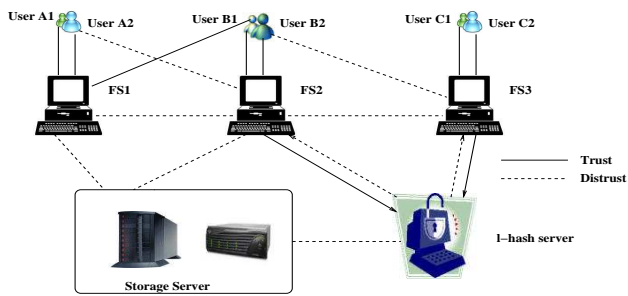}
\caption{Threat Model}
\label{fig:tmodel}
\end{figure}

Figure \ref{fig:tmodel} illustrates an instance of this threat model.
Users $A1$ and $A2$ trust the fileserver $FS1$, users $B1$ and $B2$
trust the fileserver $FS2$ and users $C1$ and $C2$ trust the
fileserver $FS3$.  User $B1$ apart from trusting $FS2$ also trusts
$FS1$.

If we consider trust domains\footnote{If we treat the entities in the system as
nodes of a graph and an edge between $i$ and $j$, if $i$ trusts $j$, then each
connected component of the graph forms a trust domain} to be made of entities
that trust each other either directly or transitively then SAFIUS guarantees
protection across trust domains. The trust relationship could be limited in
some cases (sharing of few files) or it could be complete (user trusting a
fileserver). This threat model provides complete freedom of administering the
fileservers independently and hence eases the manageability.

\section{Architecture}
\begin{figure}[t]
\includegraphics [width=3.25in, height=1.5in]{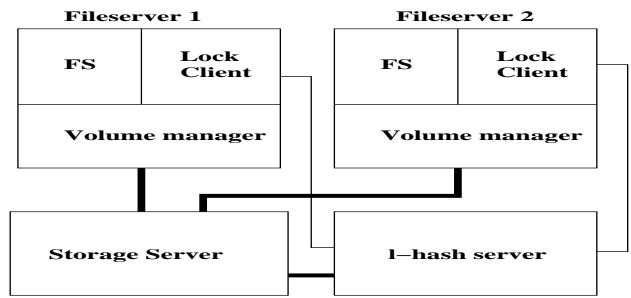}
\caption{SAFIUS Architecture}
\label{fig:architecture}
\end{figure}

The block diagram of the SAFIUS architecture is shown in figure \ref{fig:architecture}. Every fileserver in the system has a filesystem module that provides the VFS interface to the applications, a volume manager through which the filesystem talks to the storage server and a lock client module that interacts with the
l-hash server for obtaining, releasing, upgrading, or downgrading of locks. The
l-hash server, apart from serving lock requests, also distributes the hash of
inodes. The l-hash server also has the filesystem module, volume manager module
and a specialized version of lock client module and can be used like any other
fileserver in the system. The lock client modules do not interact directly
among each other, as they do not have mutual trust. The lock clients interact
transitively through the l-hash server which validates the requests. The
fileservers can  fetch the hash of inode from the trusted l-hash server and
hence fetch any file block with integrity guarantees. In figure
\ref{fig:architecture}, the thick lines represent  bulk data path and the thin
lines the metadata path. This model honours the trust assumptions stated
earlier and can scale well because each fileserver talks to the block
storage server directly.

\subsection {Block addresses, filegroups and inode-table in SAFIUS}
The blocks are addressed by their \emph{content hashes} similar to 
systems like SFS-RO \cite{sfs-ro}. It gives a \emph{write-once} 
property and blocks cannot be overwritten without changing the pointer
to the block. SAFIUS currently uses SHA-1 as the content hash and assumes that SHA-1 
collisions {\emph do not} happen. SAFIUS uses the concept of filegroups \cite{plutus},
to reduce the amount of cryptographic keys that need to be maintained.
Since ``block numbers'' are content-hashes, fetching the correct inode block
would ensure that the file data is correct. SAFIUS guarantees the integrity
of the inode block by storing the hash of the inode block in an inode hash table
 \emph{i-tbl}. Each tuple of i-tbl is called as \emph{idata}, and consists
of the inode's hash and an incarnation number. \emph{i-tbl} is stored in 
the untrusted storage server; its integrity is guaranteed by storing the hash
of the i-tbl's inode block in a local stable storage in the l-hash server.

\subsection{On-Disk structures: Inode and directory entry}
\label{on-disk-struct}

{\textbf{Inode}} The inode of a file contains pointers to data blocks either
directly or through multiple levels of indirection, apart from other meta
information found in standard UNIX filesystems. The block pointers are  SHA-1
hashes of the blocks. These apart, it also contains a 4 byte filegroup id, that
points to relevant key information to encrypt/decrypt the blocks of this file.
  
The hash of an inode corresponds to the current version of the file. If a file
is updated, then one of its leaf data block changes and hence its intermediate
metablocks (as it has a pointer to this leaf block) and ultimately the inode
block changes (this is similar to what happens in some log structured
filesystems, where writes are not done in-place, like wafl \cite{wafl}). Thus,
updating a file can be seen as moving from one version of the file to another,
with the version switch happening at a point in time when the file's idata is
updated in the i-tbl.

{\textbf{Directory entry}} Directory entries in SAFIUS are similar to the
directory entries in traditional filesystems. They contain a name and the
inode number corresponding to the name.

\subsection{Storage Server}
The granularity at which the storage server serves data is variable
sized blocks.  The storage server supports three basic operations,
namely \emph{load}, \emph{store} and \emph{free} of blocks.  A block
can be \emph{stored} multiple times, i.e. clients can issue any number
of \emph{store} requests to the same block and the block has to be
\emph{freed} that many times before the physical block can be reused at the 
server.
To prevent one user from freeing a block belonging to another user,
the storage server maintains a per \emph{inode number} reference count
on each of the stored blocks.  Each block contains a list of inode
numbers and their reference counts.  Architectures like SUNDR \cite{sundr} 
maintain a
per user reference count for the blocks.  Having a per user reference
count \emph{decimates} the possibility of seamless sharing which is
one of our design goals.  For write sharing a file between two
users $A$ and $B$ in SUNDR, the users $A$ and $B$ must belong to a
group $G$ and the file is write shareable in the group $G$. This group
$G$ has to be created and its public key need to be distributed by a
trusted entity. This restriction is due to per user reference count on
the blocks and a per user table mapping inode numbers to their hashes.
Let users $A$ and $B$ write share a file $f$ in SUNDR. If $B$ modifies
a block $k$ \emph{stored} by $A$ earlier to $l$, then $B$ cannot free
$k$, as it had not \emph{stored} it.  SAFIUS has a per inode reference
count on blocks, and the storage server does the necessary access
control to the reference count updates by looking up the filegroup
information for sharing information.  The trusted l-hash server
ratifies the access control enforced by the storage server.

The storage server authenticates the user (through public key
mechanisms) who performs the  store or free operation.  If the
uid (of the user) performing the store or free operation is same as the 
owner of the inode,
then the operation is valid. If this is not the case, then the storage
server has to verify if the current uid has enough permissions to
write to the file. If this check is not enforced, then an arbitrary
user can free the blocks belonging to files for which he has no write
access. The storage server achieves this by maintaining a cache of inode
numbers and their corresponding filegroup ids. This cache is populated
and maintained with the help of storage server. This enables seamless write 
sharing in SAFIUS.

\subsection{l-hash server: i-tbl, filegroup tree}
The l-hash server provides the basic locking service, stores
and distributes the idata of inodes to/from the itbl\footnote{i-tbl is the
persistent table indexed by inode number and contains the i-data corresponding
to an inode}. 
The l-hash server also maintains a map of inode number to
filegroup id information in a  \emph{fgrp}\footnote{It can be
realized as a file in SAFIUS} (filegroup)
tree that contains the filegroup data in the leaf blocks of the tree.
In addition, there is a persistent 64 bit monotonically increasing \emph{fgrp
  incarnation number}, a global count that indicates the number of
changes made to file sharing attributes.  The root of the filegroup
Merkle tree and the fgrp incarnation number are stored locally in the
l-hash server.  The root of the \emph{fgrp tree} is hashed with
the fgrp incarnation number to get \emph{fgrp hash}. 

\subsection{Volume Manager}
\label{sec:vm}
The volume manager
does the job of translating the read, write, and block free requests
from the fileserver to load, store or free operations that can be
issued to the storage server. The volume manager exports the
standard block interface to the filesystem module, but expects the
filesystem module to pass some additional information like hash of an
existing block (for reading and freeing) and filegroup id of the block
(for encryption or decryption).

\subsection{Encryption and hashing}
The blocks are \emph{decrypted} and \emph{encrypted}, as they enter and leave
the fileserver machine respectively by the volume manager. On a write request
the block is encrypted, hashed and stored. On a read request the blocks are
fetched from the storage server, checked for integrity (by comparing the block's
hash with that of its pointer) and is decrypted and handed over to the upper
layer. The choice of doing the encryption and decryption at the volume manager
layer was done to simplify the filesystem implementation and for performance
reasons; to delay encryption and avoid repeated decryptions on the same block.

\begin{figure}[t]
\includegraphics [width=3in, height=1.5in]{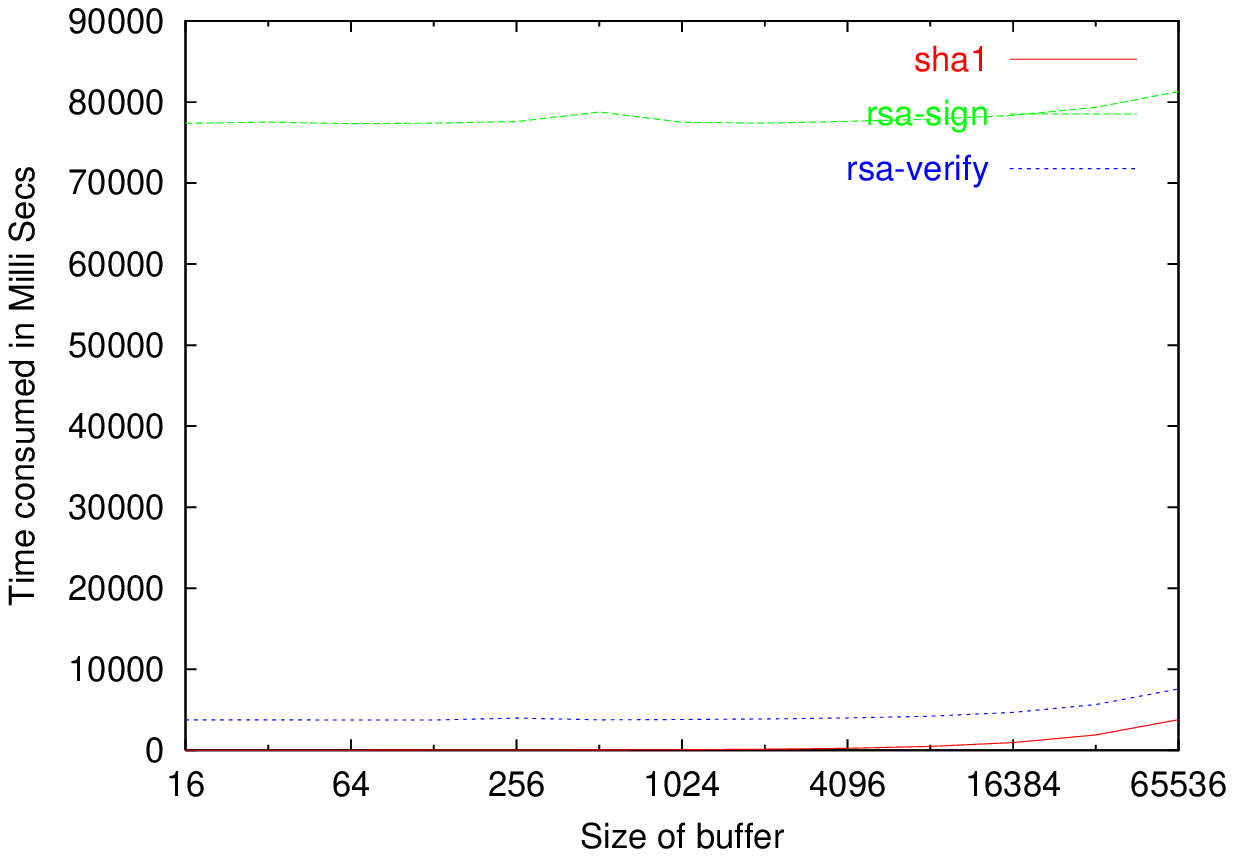}
\caption{RSA signing over various block sizes}
\label{fig:rsa-nos}
\end{figure}

\subsection{Need for non-repudiation}

Since the volume manager and the storage server are mutually
distrustful, we need to protect them from the other
party's malicious actions:
 \begin{enumerate}
 \item \textbf{Load Misses:}
   The volume manager requests a block to be a loaded but the storage
   server replies back saying that the block is not found. It could be that the
   fileserver is lying (did not store the block at all) or the storage server 
   is lying.
   
 \item \textbf{Unsolicited stores:} A block would be accounted in a
   particular user's quota, but the user can claim that he never
   stored the block.
 \end{enumerate}
 
 The first case is more serious as there is  potential data loss.
 Load operations do not alter the state of stored data and the fileserver
 would require the necessary key to decrypt it. However, for obvious reasons, 
 both store and free operations have to be non-repudiating. We achieve this 
 by tagging each of store and free operation with a signature.
 
Let
\begin{math}
  D_{u} = \{blknum,ino,uid,op,nonce,count\}.
\end{math}
For a store or free operation, the volume manager sends
\begin{math}
  \{D_u,\{D_u\}_{K_u^{-1}}\}
\end{math}
as the signature. Here \emph{blknum}
refers to the hash of the block that is to be stored or freed,
\emph{ino} refers to the inode number to which the block belongs
\emph{uid} refers to the user id of the user who is performing the
operation, \emph{nonce} is a random number that is unique across
sessions and is established with the storage server, \emph{count} is
the count of the current operation in this session and ${K_u}^{-1}$ is
the private key of the user. \begin{math} \{D_u\}_{K_u^{-1}}
\end{math} is $D_u$ signed by the private key of the user. 
The count is incremented on every store or free operation. The nonce
distinguishes two stores or frees to the same block which happens in
two different sessions, while count distinguishes two stores or frees
to the same block in the same session, hence allowing any number of 
retransmissions.  This signature captures the
current state of the operation in the volume manager. The signature is
referred to as \emph{request signature}.

\begin{figure}[t]
\includegraphics [width=3.5in, height=1.5in]{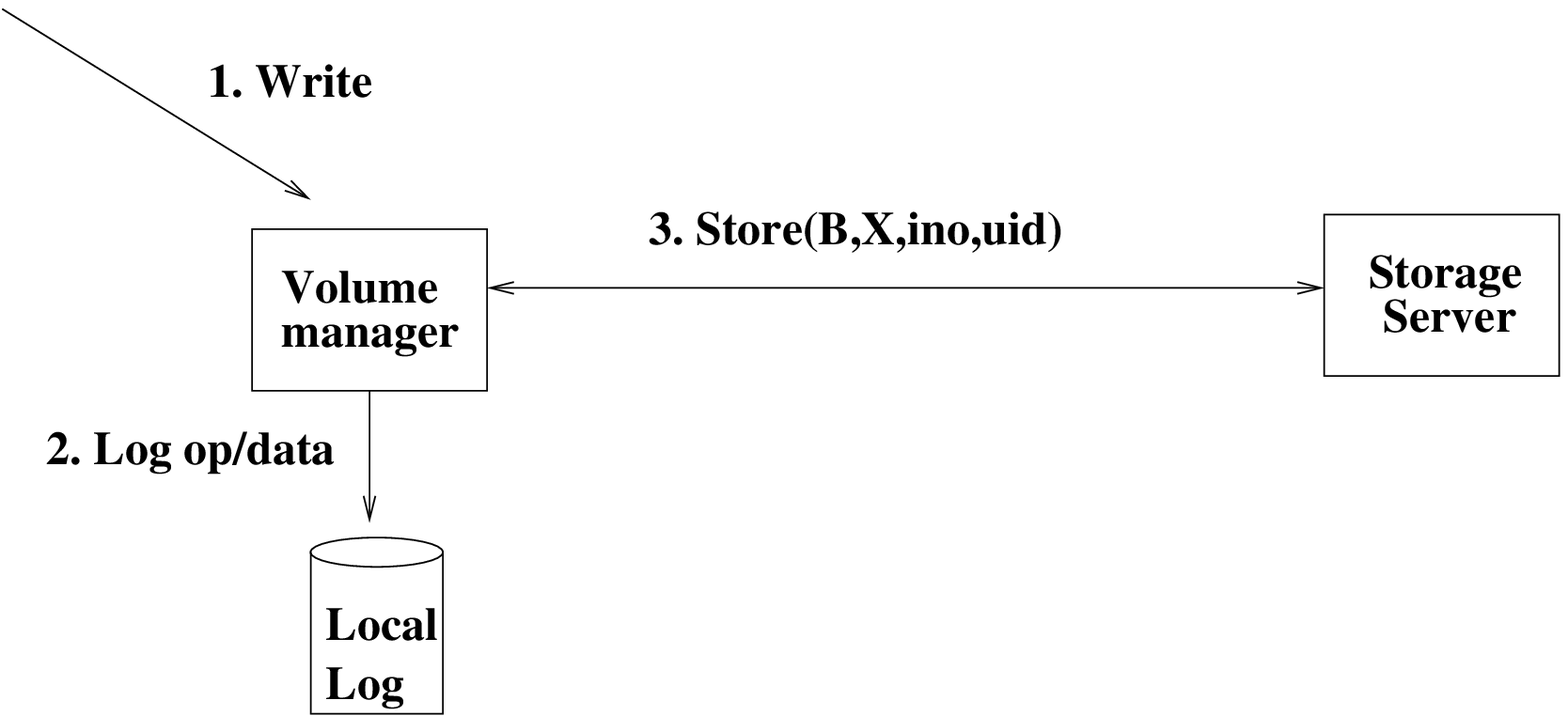}
\caption{Sequence of events on a Write}
\label{fig:write-prot}
\end{figure}

The storage server receives the signature, validates the count, uid
and nonce, and verifies the signature.  It follows the protocol
described earlier to store or free the block. On a successful
operation, it prepares and sends a reply signature. Let
\begin{math}
  D_{ss} = \{D_u,fgrphash\}.
\end{math}
The storage server sends \begin{math}
  \{D_{ss},\{D_{ss}\}_{{K_{ss}}^{-1}}\}\end{math} to the volume
manager. $D_u$ is the same as what the storage server received from
the volume manager. ${K_{ss}}^{-1}$ refers to the private key of the storage
server. The volume manager verifies that $D_u$ it receives from the
storage server is same as the one it had sent and verifies the
signature.  The signature returned to the volume manager is referred
as the \emph{grant signature}.

The grant signature prevents the storage server from denying the
stores made to a block, and in case there were free operations on the
block that resulted in the block being removed, the \emph{request signature}
of the free operation would defend the storage server. Unsolicited
stores are eliminated as the storage server will not have
request signatures for those blocks. Hence assuming that that the RSA
signatures are not forgeable, the protocol achieves non-repudiation
and hence provides perfect accountability in SAFIUS.

\subsection{Asynchronous signing}
The protocol described above has a huge performance overhead:
two signature generations and two verifications in the path of a
store or a free operation. Since signature is generated on the hash of
a block rather than the block itself, the time taken for actual signature generation, up to a
certain block size,
masks the time taken for generating the SHA-1-hash of the block.
Figure \ref{fig:rsa-nos} illustrates this. The amount of time taken
for signing a 32 byte block and the amount of time taken for signing a
16KB are comparable. However with higher block sizes, the SHA-1 cost
shows up and the signature generation cost increases linearly, as can be seen
for block sizes bigger than 16KB.

Instead of signing every operation, the protocol
signs groups of operations. The store and the free operations do not
have signing or verification in their code path and the cost of the
signing is amortized among number of store and free operations. Let
\begin{math}
  B_{u} = \{blknum,ino,op,nonce,count\},
\end{math}
The fields in this structure are same as that was in $D_u$. 
\emph{uid} field is not included
in $B_u$ as we group only operations belonging to a particular user
together and it is specified in a header for the signature block.
After a threshold number of operations or after a timeout by default, the 
volume manager packs blocks of
${B_u}$s\footnote{It contains blocks \begin{math}B_u, {B_u}', {B_u}''\end{math}} in to a block $BD_u$. The
block $BD_u$ has a header $H_u$,
\begin{math}
  H_u = \{uid,count\}
\end{math} with
{\it uid} referring to the uid of the user whose operations are being currently
bunched and signed, and {\it count} the number of operations in the
current set. Let $BD_u$ be defined as
\begin{math}
  \{H_u,B_u,{B_u}',{B_u}'',{B_u}''',..\}.
\end{math}
$BD_u$ is signed with the user's private key and
\begin{math}
  \{BD_u,\{BD_u\}_{{K_u}^{-1}}\}
\end{math}
is sent to the storage server as the \emph{request signature}. The storage
server verifies that these operations specified by $B_u$s are all
valid (they did happen) and verifies the signature on $BD_u$.  If the
operations are valid, then the storage server generates a block
$BD_{ss}$, where
\begin{math}
  BD_{ss} = \{BD_u,fgrphash\},
\end{math}
and signs it using its private key.
It sends \begin{math}
  \{BD_{ss},\{BD_{ss}\}_{{K_{ss}}^{-1}}\}\end{math} to the volume
manager which verifies that $BD_u$ is same as the one it had sent in
the \emph{request signature} and then verifies the signature.  It can be
easily seen that signing of bunch of operations is equivalent to
signing of each of these operations, provided there are no SHA-1
collisions and hence achieves non-repudiation.

\begin{figure}[t]
\includegraphics [width=3.5in, height=1.5in]{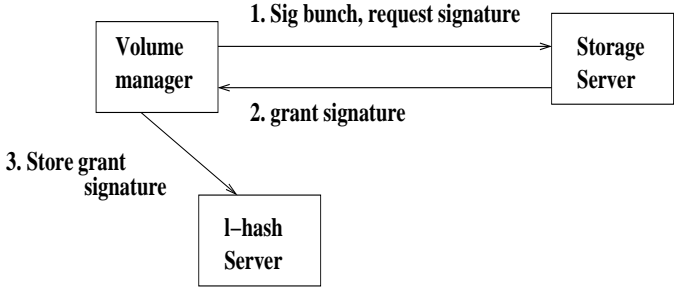}
\caption{Signature exchange}
\label{fig:sig-proto-persistent}
\end{figure}

\subsection{Need for logging}
When the operations are synchronously signed, once the store or
free operation completes, the operation cannot be repudiated by the
storage server.  But with asynchronous signing, a write or free
operation could return before the grant signature is received.  If the
storage server refuses to send the grant signature or if it fails, the
fileserver may have to retry the operations and may also have to
repeat the process of exchanging request signature for grant
signature. We need a \emph{local log} in the fileserver, where
details of the current operation are logged. We log the data for a
store request so that the store operation can be retried under error
conditions.  So, before a store or free request is sent to the server,
we log $B_u$ and the uid.  If the operation is a store operation, we
additionally log the data too.  Once the grant signature is obtained
for the bunch, the log entries can be freed. Figure
\ref{fig:write-prot} illustrates the sequence of events that happen on
a write.

\subsection{Persistence of signatures}
The request and the grant signatures should be persistent. If it were
not persistent, we cannot identify which entity violated the protocol.
The volume manager sends the grant signature
and $BD_{ss}$ to the l-hash server. It is the responsibility of the
l-hash server to preserve the signature. Figure
\ref{fig:sig-proto-persistent} illustrates this.

\begin{figure}[t]
\includegraphics [width=3.5in, height=1.5in]{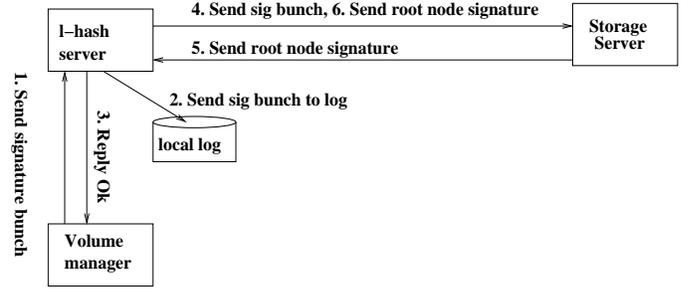}
\caption{Signature pruning protocol}
\label{fig:sig-prune-prot}
\end{figure}

\subsection{Signature pruning}
The signatures that need to be preserved at the l-hash server are on a per 
operation basis. However, the number of signatures generated is proportional to number of
store/free requests processed. A malicious fileserver can repeatedly
do a store and free to the same block to increase the number of signatures
generated. It is not possible to store all these signatures as is.
SAFIUS has a  \emph{pruning} protocol, executed between the l-hash server 
and the storage server to ensure that the amount of
space required to provide accountability is a constant function of number of
blocks used, rather than the number of operations.

In this pruning protocol, the storage server and the l-hash server agree 
upon per uid reference counts\footnote{this is different from the per inode 
reference count maintained by the storage server} on every stored block in 
the system. 
A store would increase the reference count and free would decrement it.
Each of these references must have an associated request and grant signature pair.
If the l-hash server and the storage server \emph{agree} on these reference
counts, then we can safely discard all the request and grant signatures
corresponding to this block. To achieve this, we maintain  a 
\emph{refcnt tree}, a Merkle tree, in both the l-hash server
and the storage server. The leaf blocks of this refcnt tree store
the block numbers  and the per uid reference  counts associated
with the block (only if at least one reference count is non-zero).
If the root block of the \emph{refcnt tree} is same in both the l-hash server
and storage server, then both parties must have the same reference counts on
the leaf blocks. Figure \ref{fig:sig-prune-prot} illustrates the signature pruning protocol.
Since the storage server has signed the root of the tree that it had
generated, there cannot be a load miss for a valid block from the storage
server side. The \emph{refcnt tree} in the l-hash server helps provide   
accountability. To save some space,
the \emph{refcnt tree} does not store the reference count map for all
blocks. It has a table of unique reference count entries (mostly blocks
owned by one user only) and the \emph{refcnt tree}'s leaf blocks merely
have  a pointer to this table.

\subsection{The filesystem module}
\label{sec:fs}
The filesystem module provides the standard UNIX like interface for
the applications, so that applications need not be re-written.
However, owing to its relaxed threat model, the file system has the 
following restrictions:

\begin{itemize}
  
\item Distributed filesystems like frangipani \cite{frangipani} and GFS
\cite{gfs} have a notion of a per node log, which is in a universally
accessible location. Any node in the cluster can replay the log. In the threat
model that we have chosen, the fileservers do not trust each other; so it  is
not possible for one fileserver to replay the log of another fileserver to
restore filesystem consistency.
  
\item Traditional filesystems have a notion of consistency in which each block
in the system is in use or is free. In case of SAFIUS, this notion of
consistency is tough to achieve.

\item Given our relaxed threat model, it is the responsibility of the
fileservers to honour the filesystem structures. If they do not, there is a
possibility of filesystem inconsistency. However, the SAFIUS architecture
guarantees complete isolation of the effects of the misbehaving entity to its
own trust domain.

\end{itemize}

\subsection{Read/Write control flow}
The fileservers get the root directory inode's idata during mount time.
Subsequent file or directory lookups are done in the same way as 
in a standard UNIX filesystems.

{\textbf{Reads:}}
The inode's idata  fetched from the l-hash server is the
only piece of metadata that the fileserver needs to obtain from the l-hash
server. The filesystem module can fetch the blocks it wants from the storage
server directly by issuing a read to the volume manager. During the read call,
when the fileserver requests a shared lock on the file's inode, the l-hash
server, apart from granting the lock, also sends the idata of the inode. Using
this idata it can fetch the inode block and hence the appropriate intermediate
blocks and finally the leaf data block, which contains the offset requested.

{\textbf{Writes:}} 
Write operations from the fileserver usually proceed by first obtaining an
exclusive lock on the inode. While granting the exclusive lock, the l-hash
server also sends the latest hash of the inode as a part of idata.  After the
update of the necessary blocks including the inode block, the hash of the new
inode block corresponds to the new version of the file. As long as the idata in
the l-hash server is not updated with this, the file is still in the old
version. When the new idata corresponding to this file -- hence inode, is
updated in the i-tbl, the file moves to a new version. The l-hash makes sure
that the current user has enough permissions to update the inode's hash. Now
the old data blocks and metablocks that have been replaced by new ones in the
new version of the file have to be freed. The write is not visible to other
fileservers until the inode's idata is updated in the l-hash server.  Since
this is done before releasing the exclusive lock,  any intermediate reads to
the file would have to wait.

\subsection{Logging}
Journaling is used by filesystems to speed up the task of restoring
the consistency of the filesystem after a crash.  Many filesystems use
a redo log for logging their metadata changes. During recovery after a
crash, the log data is replayed to restore consistency. In SAFIUS,
pending updates to the filesystem during the time of crash do not
affect the consistency of the filesystem as long as the fileservers do
not free any block belonging to previous version of the file and the
inode's idata is not updated.  When the inode's idata is updated in
the l-hash server, the file moves to the next version and all
subsequent accesses will see the new version of the file.  The overwritten 
blocks have to be freed when the idata update in l-hash server is successful 
and the new blocks that were written to should be freed if the idata update
failed for some reason. SAFIUS uses an undo-only operation log to achieve this.

\begin{figure}[t]
\includegraphics [width=3.5in, height=1.5in]{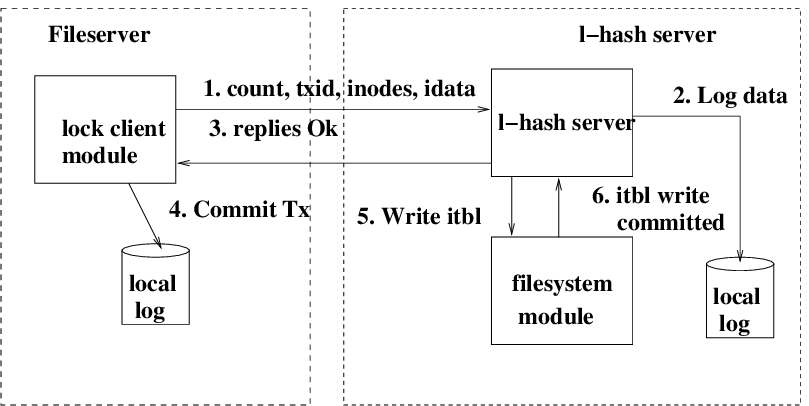}
\caption{Store inode data protocol}
\label{fig:st-inode-data}
\end{figure}

\subsection{Store inode data protocol}
To ensure that the system is consistent, the idata of an inode in the
i-tbl of the l-hash server has to be updated atomically, i.e. the
inode's idata has to be either in the old state or in the new state
and the fileserver that is performing the update should be able to
know whether the update sent to the l-hash server has succeeded or
not. Fileservers take an exclusive lock on the file when it is opened
for writing. After flushing the modified blocks of the
file and before dropping the lock it holds,
the fileserver executes the \emph{store-inode data} protocol with the
l-hash server to ensure consistency. The store inode data protocol is
begun by the fileserver sending the count and the list of inodes and
their idata to the l-hash server.  The l-hash server stores inodes' new idata
in the i-tbl atomically (either all of these inodes' hashes are updated or none
of them are updated), employing a local log. It  also \emph{remembers} the last
txid received as a part of \emph{store inode data} protocol from each
fileserver.

After receiving a reply from the l-hash server, the fileserver writes
a commit record to the log and commits the transaction, after which
the blocks that are to be freed are queued for freeing and log space
is reclaimed. The l-hash server \emph{remembers} the latest txid from
the fileservers to help the fileservers know if their last execution
of \emph{store inode data} had succeeded. If the fileserver had
crashed immediately after sending the inode's hash, it has no way of
knowing whether the l-hash server received the data and had updated
the i-tbl. If it had updated the i-tbl, the transaction has to be
committed and the blocks meant for freeing need to be freed. If it is not
the case, then the transaction has to be aborted and the blocks
written as a part of that transaction have to be freed.

On recovery, the fileserver contacts the l-hash server to get the
last txid that had updated the i-tbl. If that txid does not have a
commit record in the log, then the commit record is added now and the
recovery procedure is started. Since all the calls to \emph{store inode data}
protocol are serialized within a fileserver and the global changes are
visible only on updates to the i-tbl, this protocol will ensure
consistency of the filesystem. The store inode data protocol takes a
list of $\langle$inode number, idata$\rangle$ pair instead of a single
inode number, idata pair. This is to ensure that dependent inodes are
flushed atomically.  For instance, this is useful during file creation and 
deletion, when the file inode is dependent on the directory inode.


\subsection{File Creation}
File creation involves obtaining a free inode number, creating a
new disk inode and updating the directory entry of the parent
directory with the new name-to-inode mapping. Inode numbers
are generated on the fileservers autonomously without consulting any
external entity. Each fileserver stores a persistent
bitmap of free local inode numbers locally.  This map is updated after
an inode number is allocated for a new file or directory.  

\subsection{File Deletion}
Traditional UNIX systems provide a \emph{delete on close} scheme for unlinks.
To provide similar semantics in a distributed filesystem, one has to
keep track of open references to a file from all the nodes and the
file is deleted by the last process which closes the file, among all the
nodes. This warrants that we need to maintain some global information
regarding the open references to files.  In a
NFS like environment, where the server is stateless, reading and
writing to a file that is unlinked from some other node results in a
\emph{stale file handle} error.
SAFIUS' threat model does not permit similar unlink semantics. 
So we define a \emph{simplified unlink semantics} for file deletes in
SAFIUS. Unlink in SAFIUS removes the directory entry and decrements
the inode reference count, but it defers deletion of the file as long
as any process in the \emph{same node}, from which unlink was called,
has an open reference to the file. The last process on the node, from
which unlink was called, deletes the file. Subsequent reads and writes
from other nodes to the file do not succeed and return stale file handle
error. There
would not be any new reads and writes to the file from the node that
called unlink as the directory entry is removed and the last process
that had an open reference has closed the file. This semantics honours
the standard \emph{read after write} consistency. As long as the file
is not deleted, a read call following a write returns the latest
contents of the file.  After a file is deleted, subsequent reads and 
writes to the file do not succeed, and hence \emph{read after write} 
consistency.

\begin{figure}[t]
\includegraphics [width=3.5in, height=2.0in]{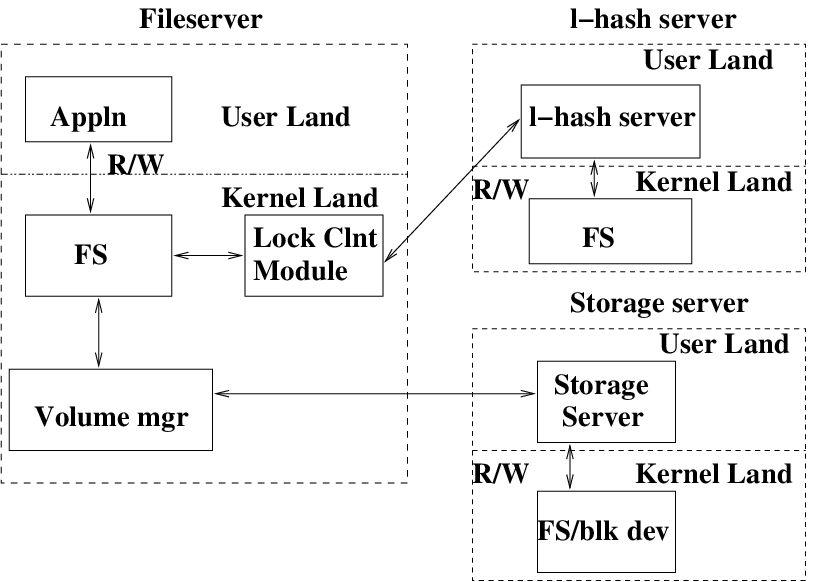}
\caption{Implementation Overview}
\label{fig:impl-overview}
\end{figure}

The inode numbers have to be freed for re-allocation. As mentioned
earlier, inode numbers identify the user and the machine id who owns
the file. If the node which unlinks the file is same as the one which
has created it, then the inode number can be marked as free in the
allocation bitmap. But if unlink happens in another machine, then the
fact that the inode number is free has to be communicated to that
machine. Since our threat model does not assume two fileservers to
trust each other, the information has to be routed through the l-hash
server. The l-hash server sends the \emph{freed inode numbers} list to
the appropriate fileserver, during mount time (when the
fileserver fetches the root directory inode's idata).

\subsection{Locking in SAFIUS}
SAFIUS uses the Memexp protocol of OpenGFS \cite{gfs} with some minor
modifications.  The l-hash server ensures that the
current uid has enough permissions to acquire the lock in the
particular mode requested.  The lock numbers and the inode numbers
have a one to one correspondence and hence we can derive the inode
number from the lock number. Using the lock number, the l-hash server
obtains the filegroup id and hence the permissions.

OpenGFS has a mechanism of callbacks wherein a node that needs a
lock, currently held by another node, sends a  message to that
node's callback port. The node which holds the lock downgrades the
lock if the lock is not in use. In SAFIUS, since the callback cannot
be directly sent (the two fileservers would be mutually distrusting),
the callbacks are routed through the l-hash server. 

\subsection{Lock client module}
\label{sec:lclient}
The lock module in the fileserver handles all the client side
activities of the lock protocol that was briefly described in the
previous section. Apart from this, it also does the job of fetching
the idata corresponding to an inode number from the l-hash server.  It
also executes the store inode data protocol with the l-hash server to
ensure atomic updates of list of inodes and their idata.  
Figure \ref{fig:st-inode-data} illustrates the protocol.
The protocol guarantees atomicity of updates to a set of inodes and their
idata.

\section{Implementation}
SAFIUS is implemented in the GNU/Linux environment. Figure
\ref{fig:impl-overview} depicts the various modules in SAFIUS and
their interaction.  The base code used for the filesystem and lock
server is OpenGFS-0.2\footnote{http://opengfs.sourceforge.net/} and base
code used for the volume manager is GNBD-0.0.91. The Memexp lock server in 
OpenGFS was modified to be the l-hash server to manage locks and to store 
and distribute idata of the inodes. The volume
manager, the filesystem module and the lock client module reside
inside the kernel space, while the storage server and l-hash server
are implemented as user space processes. The current implementation of
SAFIUS does not have any key management scheme and keys are manually
distributed. The itbl has to reside in the untrusted
storage as it has to hold the idata for all the inodes in the system. 
Consequently, the itbl's integrity and freshness has to be guaranteed.  
We achieve this by storing the itbl information in a special file in 
the root directory of the filesystem (.itbl). The l-hash server stores
the idata of this file in its local stable storage. The idata of the itbl
serve as the bootstrap point for validating any file in this filesystem.

\begin{figure}[t]
\includegraphics[height=2.5in, width=4in] {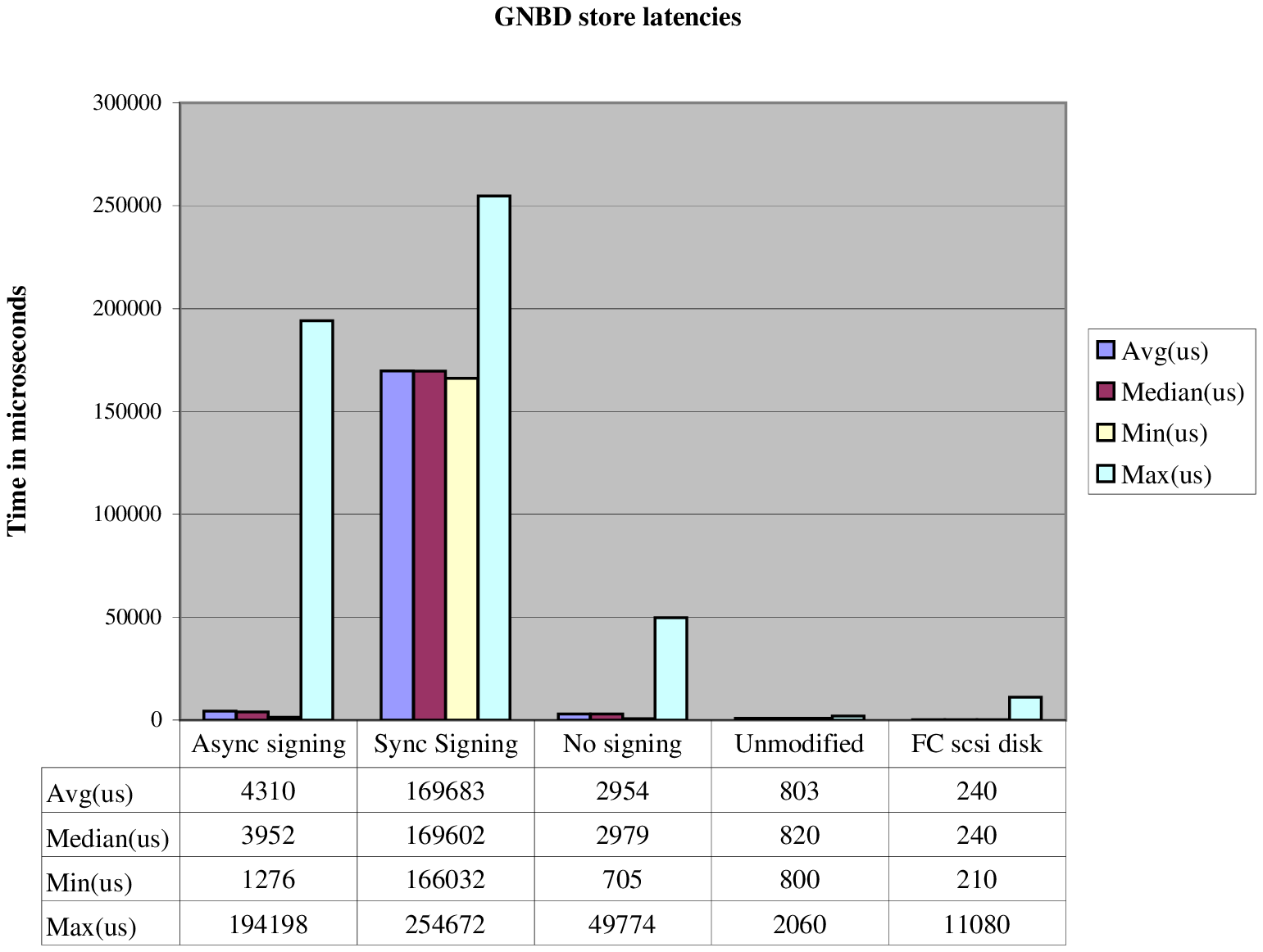}
\caption{Store latencies}
\label{fig:gnbd}
\end{figure}

\section{Evaluation}

The performance of SAFIUS has been evaluated with the following hardware
setup. The fileserver is a Pentium III 1266 MHz machine with 896MB of
physical memory. A machine with a similar configuration serves as the
l-hash server, when the l-hash server and fileserver are different.  A
Pentium IV 1.8 GHz machine with 896MB of physical memory functions as
storage server. The storage server is on a Gigabit Ethernet and the
fileserver and l-hash servers are on 100Mbps Ethernet. The fileserver
and l-hash servers have a log space of 700MB on a fibre channel SCSI
disk.  The storage server uses a file on the ext3 filesystem (over a
partition on IDE hard disk) as its store. The fileserver, storage
server and the l-hash server all run Linux kernel. The
performance numbers of SAFIUS are reported in comparison with an OpenGFS setup. A GNBD device served as the shared block device.  For
the OpenGFS experiments, the storage server machine hosts the GNBD
server and the fileserver machine hosts the OpenGFS FS client.  The
l-hash server machine runs the memexp lock server of OpenGFS.

\begin{figure*}
\begin{tabular}{l}

\begin{minipage}[c]{2.3in}
\includegraphics [width=0.85\linewidth]{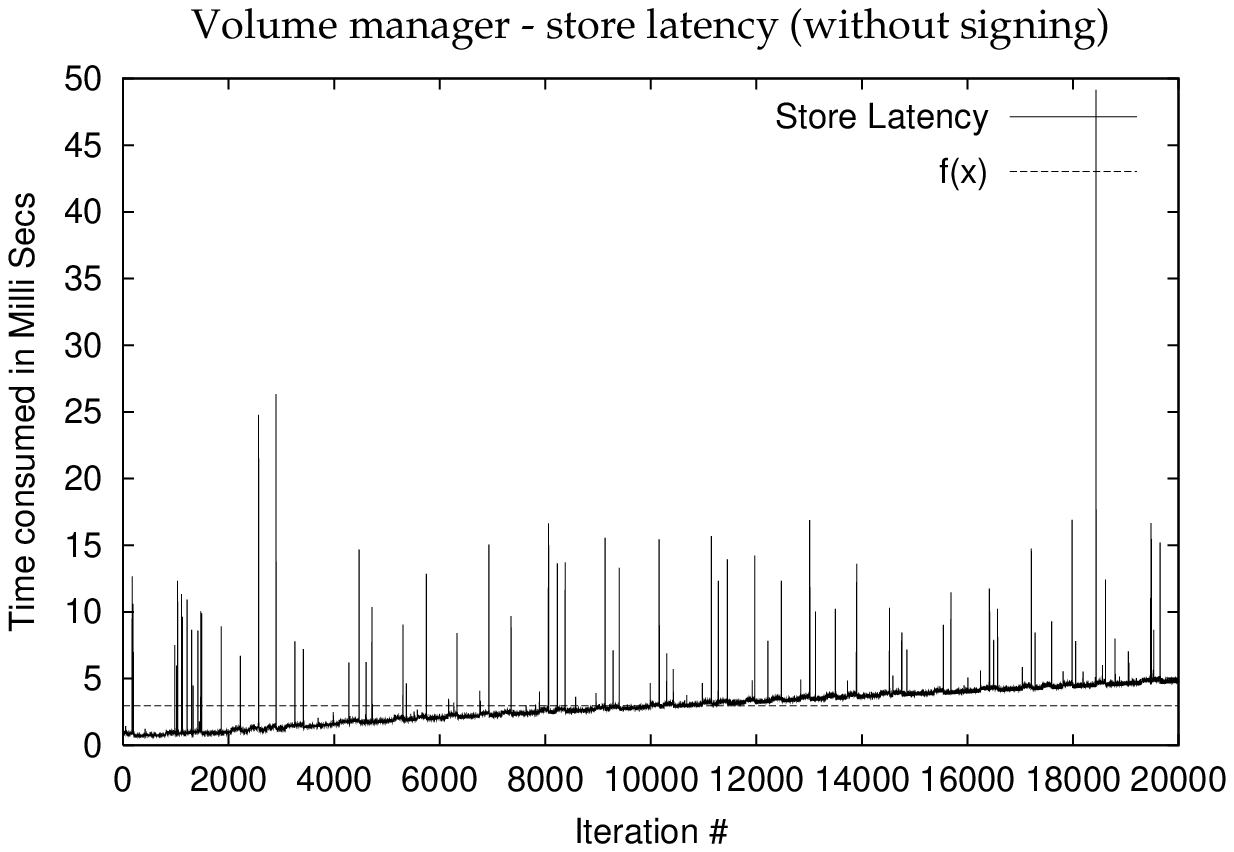}\caption{No-signing}
\label{fig:nsign}
\end{minipage}

\begin{minipage}[c]{2.3in}
\includegraphics [width=0.85\linewidth]{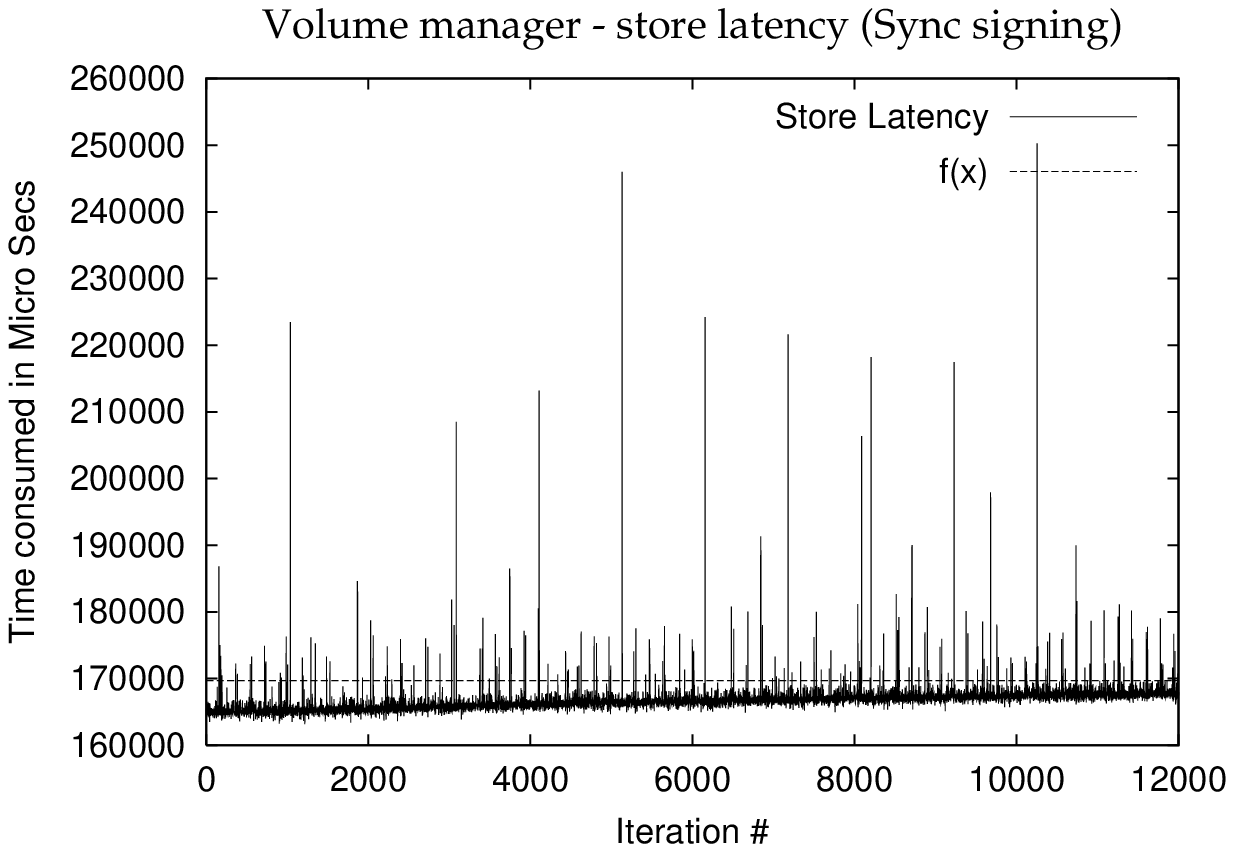}\caption{Sync-signing}
\label{fig:ssync}
\end{minipage}

\begin{minipage}[c]{2.3in}
\includegraphics [width=0.85\linewidth]{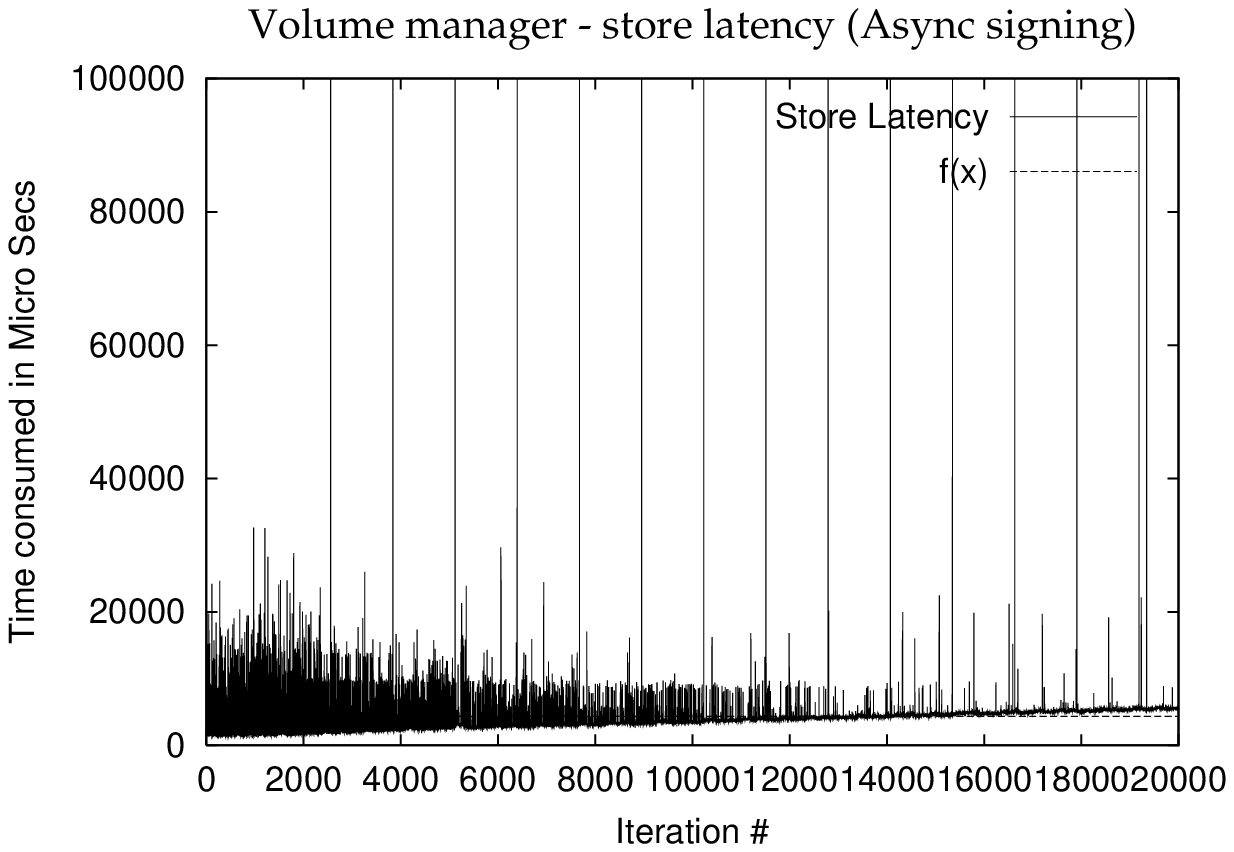}\caption{Async-signing}
\label{fig:async}
\end{minipage}

\end{tabular}
\end{figure*}

Basically two sets of configuration are studied: one in which the
l-hash server and the fileserver are the same machine and another in
which the l-hash server and the fileserver are two different machines.
In OpenGFS setup the two configurations are: one in which the lock
server and the FS client were in the same machine and another in which
they were in two physically different machines. A thread,
\emph{gfs\_glockd} in SAFIUS, wakes up periodically to drop the unused
locks.  The interval in which this thread is kicked in is used as a
parameter of study. The performance of 
SAFIUS configurations are reported with and
without encryption/ decryption.  Hence, the performance numbers
reported are for eight different combinations for SAFIUS and two
different combinations for OpenGFS.

\begin{enumerate}
\item {\textbf{SAFIUS-I30:} It is a SAFIUS setup in which the l-hash
    server and the fileserver are the same machine. The gfs\_glockd
    interval is 30 seconds.}
\item {\textbf{SAFIUS-I10:} It is a SAFIUS setup in which the l-hash
    server and the fileserver are the same machine. The gfs\_glockd
    interval is 10 seconds.}
\item {\textbf{SAFIUS-D30:} It is a SAFIUS setup in which the l-hash
    server and the fileserver are different machines. The gfs\_glockd
    interval is 30 seconds }
\item {\textbf{SAFIUS-D10:} It is a SAFIUS setup in which the l-hash
    server and the fileserver are different machines. The gfs\_glockd
    interval is 10 seconds }
\item {\textbf{OpenGFS-I:} It is an OpenGFS setup in which the memexp
    lock server and the filesystem run on the same machine}
\item {\textbf{OpenGFS-D:} It is an OpenGFS setup in which the memexp
    lock server and the filesystem run on different machines}

\end{enumerate}

SAFIUS-I30E, SAFIUS-I10E, SAFIUS-D30E, SAFIUS-D10E are the SAFIUS
configurations with encryption/decryption.

\subsection{Performance of Volume manager - Microbenchmark }

As described in Section \ref{sec:design}, SAFIUS volume manager uses
asynchronous signing to avoid \emph{signing} and \emph{verification}
in the common store and free path. 
Experiments have been conducted to measure the latencies of load and store
operations. An ioctl interface in the volume manager code is used for
performing loads and stores from the userland, bypassing the buffer cache
of the kernel. A sequence of 20000 store operations are performed
with the data from /dev/urandom. The fileserver machine is used for
issuing the stores.  Figure \ref{fig:async} shows the plot of the
latency in Y-axis and the store operation sequence number along
X-axis.  Approximately once every 1000 operations, there is a huge
vertical line, signifying a latency of more than 100ms \footnote{The Y-axis scale is trimmed, the latencies are about 170ms}. This is when
the signer thread kicks in to perform the signing. As observed in
section \ref{sec:design}, the latency of a signing operation is about
80ms on a Pentium IV machine.  As can be observed
from figure \ref{fig:rsa-nos}, the cost of signing is constant till
the block size is 32KB and grows linearly after that. This increase
corresponds to the cost of SHA-1, which is amortized by the RSA
exponentiation cost for smaller block sizes.  The plot in figure
\ref{fig:async} also shows the mean and median of the latencies
measured. We also have studied GNBD store latencies for various combinations.
The results are reported in Fig. \ref{fig:gnbd}. As to be expected, synchronous
signing incurs the highest overhead while the asynchronous signing, on an
average, appears to be only 30\% costlier compared to no-signing.

The next subsection describes the performance studies conducted for the
filesystem module.

\subsection{Filesystem Performance - Microbenchmark and Popular benchmarks}
\begin{figure}[t]
\includegraphics[height=3in, width=3.5in] {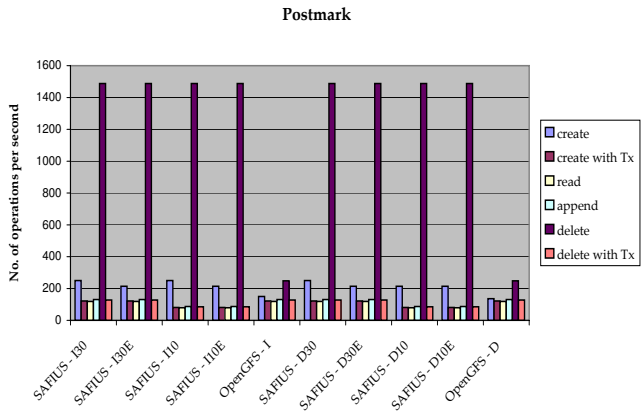}
\caption{Postmark}
\label{fig:Postmark}
\end{figure}

The performance of the filesystem module has been analyzed by running
the Postmark \footnote{Benchmark from Network Appliance}
microbenchmark suite and compilation of OpenSSH, OpenGFS and Apache
web server sources. As described in the beginning of this section,
numbers are reported for six configurations. Figure \ref{fig:Postmark}
shows the numbers obtained by running Postmark suite. The Postmark
benchmark has been run with the following configuration:

\begin{verbatim}
set size 512 10000
set number 1500
set seed 2121
set transactions 500
set subdirectories 10
set read 512
set write 512
set buffering false
\end{verbatim}

Currently, owing to the simple implementation of the storage server,
logical to physical lookups take a lot of time and hence Postmark with 
bigger configurations take a long time to complete.
Hence we report Postmark results only for the smaller configurations.
SAFIUS beats OpenGFS in metadata intensive operations like create and
delete. Creations and deletions in SAFIUS are not immediately
committed (even to log) and are committed only when the lock is
dropped and hence the explanation.  With the current input
configuration, reads and appends for SAFIUS-D30, SAFIUS-I30, OpenGFS-I
and OpenGFS-D seem to be the same.  Bigger configurations may show
some difference. SAFIUS-I10 and OpenGFS-D10 seem to perform poorly for
appends and reads due to flushing of data belonging to files that
would anyway get deleted.  There is no difference between the l-hash
server being in the same machine or in different machine for Postmark
suite. For the Postmark suite, there is not much difference between
the configurations that has encryption/decryption and the ones that
doesn't have.

\begin{figure}[t]
\includegraphics[height=2.25in, width=3.5in] {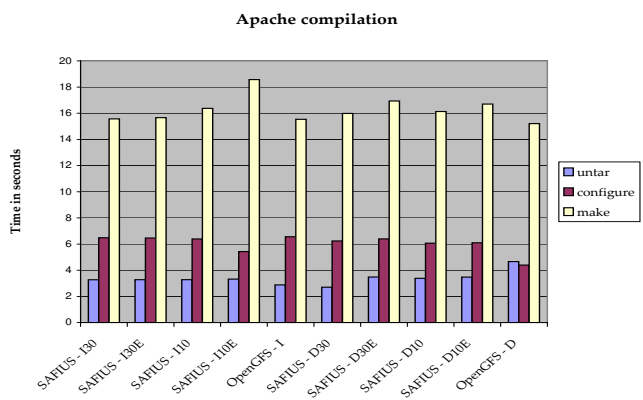}\caption{Apache compilation}
\label{fig:apache}
\end{figure}

\begin{figure}[t]
\includegraphics[height=2.25in, width=3.5in] {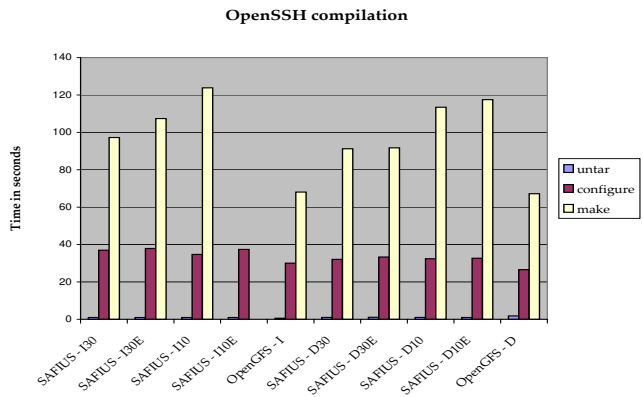}
\caption{OpenSSH compilation}
\label{fig:openssh}
\end{figure}

\begin{figure}[t]
\includegraphics[height=2.25in, width=3.5in] {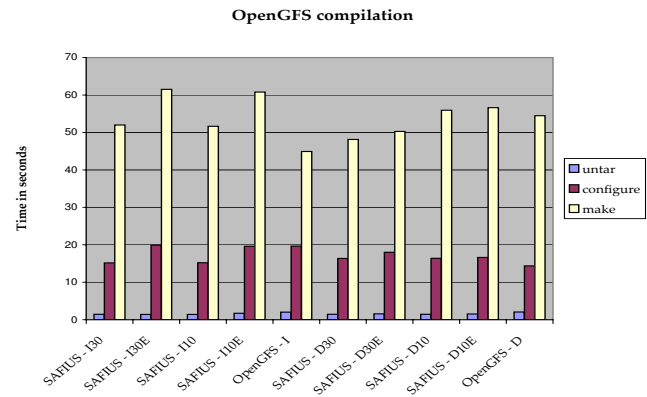}
\caption{OpenGFS compilation}
\label{fig:opengfs}
\end{figure}

Next three performance tests involved compilation of OpenSSH,
OpenGFS and Apache web server. Three activities were performed on the source
tree: untar (tar zxvf) of source, configure and make. Time taken for each of
these operations for all the ten configurations described before is reported. Figure \ref{fig:openssh} is the result of OpenSSH compilation. All SAFIUS configurations take twice the amount of time for untar compared
to OpenGFS-I, while OpenGFS-D takes four times the time taken by
OpenGFS-I for untar.  OpenSSH compilation did not complete in
SAFIUS-I10E. Time taken for make in SAFIUS-D30 and SAFIUS-D30E, seems
to be almost same, while SAFIUS-I30E takes 10\% more time than
SAFIUS-I30. The best SAFIUS configurations (SAFIUS-D30 and
SAFIUS-D30E) are within 120\% of the best OpenGFS configuration
(OpenGFS-D). The worst SAFIUS configurations (SAFIUS-I10E and
SAFIUS-D10E) are within 200\% of worst OpenGFS configuration
(OpenGFS-I).  The next performance test was compilation of Apache
source. Figure \ref{fig:apache} shows the time taken for untar,
configure and make operations of Apache source compilation. SAFIUS-D30
gives the best performance for untar and OpenGFS-D gives the worst.
This is probably because of reduced interference for syncing the idata
writes.  OpenGFS-D does the best for make and configure. Among SAFIUS
configurations, SAFIUS-D10 does the best for configure and SAFIUS-I30
does the best for make. The last performance test is OpenGFS compilation.
\emph{make} was run from the src/fs subtree instead of the toplevel source tree. Figure \ref{fig:opengfs} shows the time taken for untar, configure
and make operations for compiling OpenGFS. SAFIUS-I30E and SAFIUS-I10E takes
about 120\% of time taken by SAFIUS-I30 and SAFIUS-I10 respectively for running
configure and make. The best SAFIUS configuration for running make (SAFIUS-D30)
takes around 115\% of time taken by best OpenGFS configuration (OpenGFS-I). The
worst SAFIUS configuration for running make (SAFIUS-I30E) takes about 115\% of
time taken by worst OpenGFS configuration (OpenGFS-D).

We can conclude, from the experiments run, that SAFIUS
seems to be comparable (or sometimes better) to  OpenGFS for metadata intensive operations
and around 125\% of the best OpenGFS configuration without
encryption/decryption, and around 150\% of the best OpenGFS
configuration with encryption/decryption and for other operations.

\section{Conclusions \& Future Work }
\label{cha:conclusions}

In this work the design and implementation of a secure distributed
filesystem over untrusted storage was discussed. SAFIUS provides
confidentiality, integrity, freshness and accountability guarantees,
protecting the clients from malicious storage and the storage from
malicious clients. SAFIUS requires that trust be placed on the
lock-server (\emph{l-hash} server), to provide all the security
guarantees; a not so unrealistic threat model. For the applications,
SAFIUS is like any other filesystem; it does not require any change of
interfaces and hence has no compatibility issues. SAFIUS uses the
\emph{l-hash} server to store and retrieve the hash codes of the inode
blocks. The hash codes reside on the untrusted storage and the integrity
of the system is provided with the help of a secure local storage in the
\emph{l-hash} server. SAFIUS is flexible; users choose which client
\emph{fileservers} to trust and how long.  SAFIUS provides ease of
administration; the \emph{fileservers} can fail and recover without
affecting the \emph{consistency} of the filesystem and without the
involvement of another entity.  With some minor modifications, SAFIUS
can easily provide consistent snapshots of the filesystem (by not
deleting the overwritten blocks). The performance of SAFIUS is
promising given the security guarantees it provides.  A detailed
performance study (under heavier loads), has to be done in-order to
establish the consistency in performance.

Possible avenues for future work are:

\begin{itemize}
\item {\textbf{Fault tolerant distributed l-hash server:} The l-hash
    server in SAFIUS can become a bottleneck and prevent scalability
    of fileservers. It would be interesting to see how the system
    performs when we have a fault tolerant distributed l-hash server
    in place of existing l-hash server. The distributed lock protocol
    should work without assuming any trust between fileservers. }
  
\item {\textbf{Optimizations in storage server:} The current
    implementation of the storage server is a simple request response
    protocol that serializes all the requests. It would be a
    performance boost to do multiple operations in parallel. This may
    affect the write ordering assumptions that exist in the current
    system.}
  
\item {\textbf{Utilities:} Userland filesystem debug utilities and
    failure recovery utilities have to be written.}
  
\item {\textbf{Key management:} SAFIUS does not have a key management
    scheme and no interfaces by which the users can communicate to the
    fileservers their keys. This would be an essential element for the
    system.}

\end{itemize}

The source code for SAFIUS is available on request.

\bibliographystyle{plain}

\markright{Bibliography}
\bibliography{ref}

\begin{thebibliography}{1}

\bibitem{sfs-ro}
Kevin Fu, M.~Frans Kaashoek, and David Mazieres.
\newblock Fast and secure distributed read-only file system.
\newblock {\em Computer Systems}, 20(1):1--24, 2002.

\bibitem{wafl}
D.~Hitz, J.~Lau, and M.~Malcolm.
\newblock File system design for an {NFS} file server appliance.
\newblock In {\em Proceedings of the {USENIX} Winter 1994 Technical
  Conference}, pages 235--246, San Fransisco, CA, USA, 1994.

\bibitem{sundr}
David~Mazières Jinyuan~Li, Maxwell~Krohn and Dennis Shasha.
\newblock Secure untrusted data repository (sundr).
\newblock Technical report, NYU Department of Computer Science, 2003.

\bibitem{plutus}
M.~Kallahalla, E.~Riedel, R.~Swaminathan, Q.~Wang, and K.~Fu.
\newblock Plutus - scalable secure file sharing on untrusted storage.
\newblock In {\em In Proceedings of the Second USENIX Conference on File and
  Storage Technologies (FAST). USENIX}, March 2003.

\bibitem{ivy}
Athicha Muthitacharoen, Benjie Chen, and David Mazieres.
\newblock A low-bandwidth network file system.
\newblock In {\em Symposium on Operating Systems Principles}, pages 174--187,
  2001.

\bibitem{gfs}
Steven~R. Soltis, Thomas~M. Ruwart, and Matthew~T. O'Keefe.
\newblock The {Global File System}.
\newblock In {\em Proceedings of the Fifth {NASA} Goddard Conference on Mass
  Storage Systems}, pages 319--342, 1996.

\bibitem{pfs}
Christopher~A. Stein, John~H. Howard, and Margo~I. Seltzer.
\newblock Unifying file system protection.
\newblock In {\em In Proc. of the USENIX Technical Conference}, pages 79--90,
  2001.

\bibitem{frangipani}
Chandramohan~A. Thekkath, Timothy Mann, and Edward~K. Lee.
\newblock Frangipani: A scalable distributed file system.
\newblock In {\em Symposium on Operating Systems Principles}, pages 224--237,
  1997.

\bibitem{tdb}
R.~Vingralek U.~Maheshwari and B.~Shapiro.
\newblock How to build a trusted database system on untrusted storage.
\newblock In {\em OSDI: 4th Symposium on Operating Systems Design and
  Implementation}, 2002.

\end{thebibliography}

\end{document}